\documentclass[twocolumn,prd,nofootinbib,aps,prl,floats,floatfix,amsmath,amssymb,longbibliography,secnumarab
ic]{revtex4-1} %
\usepackage[final]{graphicx}
\usepackage{hyperref}
\usepackage{amsmath}
\usepackage{bbm}
\usepackage{amsfonts}
\usepackage{amssymb}
\usepackage{latexsym}
\usepackage{graphicx}
\usepackage[english]{babel}
\usepackage{multirow}
\usepackage{float}
\usepackage{url}
\usepackage{slashed}
\usepackage{xcolor} 
\usepackage[utf8]{inputenc}
\usepackage{verbatim}
\usepackage{stackengine}

\def\sfrac#1#2{{\textstyle{#1\over #2}}}
\newcommand{\be}{\begin{equation}}
\newcommand{\ee}{\end{equation}}
\newcommand{\ba}{\begin{array}}
\newcommand{\ea}{\end{array}}
\newcommand{\bea}{\begin{eqnarray}}
\newcommand{\eea}{\end{eqnarray}}

\newcommand{\nn}{\nonumber}

\usepackage[utf8]{inputenc}

\begin{document}
\title{Self-interacting dark baryons}

\author{James M.\ Cline}
\affiliation{McGill University, Department of Physics, 3600 University St.,
Montr\'eal, QC H3A2T8, Canada}
\author{C\'edrick Perron}
\affiliation{University of Toronto, Department of Physics,
60 St George St, Toronto, ON M5S1A7, Canada}
\begin{abstract}
Using results from lattice QCD, it is possible to quantitatively
design models of dark baryons leading to velocity-dependent
self-interaction cross sections that match the values needed for
solving small-scale structure problems of standard cold dark matter. 
However it is not obvious that the main dark matter component in such
models will be nucleons rather than large nuclei, or dark pions or
atoms, whose scattering properties would be different.  We first
identify the parameters of a dark SU(3) sector analogous to QCD---the
confinement scale $\Lambda$ and pion mass $m_\pi$---needed to
reproduce desired self-interaction cross sections. Then we show that
these values can generically be compatible with the absence of a
sufficiently stable deuteron bound state, and hence leading to no
heavier nuclei, thus establishing the consistency of the scenario for
self-interacting dark nucleons.  The range of dark photon masses
needed to avoid dominant pion or atomic dark matter is determined, as
well as allowed values for the kinetic mixing parameter.  The dark
proton might be detected directly in future searches, by dark photon
exchange. \end{abstract} 

\maketitle

\section{Introduction}
Over two decades ago, it was proposed that dark matter (DM) with strong self-interactions \cite{Spergel:1999mh,Dave:2000ar} could address a discrepancy between gravitational $N$-body simulations of galaxy formation, which
predict cuspy central density profiles \cite{Navarro:1995iw,Navarro:1996gj}, versus observations that indicate otherwise \cite{1994ApJ...427L...1F,Walker_2011,10.1093/mnras/stu474,de_Blok_2001,2002A&A...385..816D,10.1111/j.1365-2966.2004.07836.x,Newman_2013}.  Subsequent to the early observations, several other discrepancies were identified, known as the missing satellites \cite{10.1093/mnras/264.1.201} and too-big-too-fail \cite{10.1111/j.1745-3933.2011.01074.x,10.1093/mnras/stu1477} problems, that could be ameliorated by including DM self-interactions
in the numerical simulations \cite{10.1093/mnras/sts514,10.1093/mnras/sts535}. The missing satellites problem has tended to disappear over time (even becoming a ``too many satellites problem''), as observations and simulations have improved \cite{Kim:2017iwr,2019MNRAS.486.4545F,Kim:2021zzw}, but the core-cusp problem seems more robust.

The small scale structure issue is complicated by baryonic feedback, originating from supernova shock waves sweeping material out of the dense inner regions of galaxies \cite{10.1093/mnras/stz1890}, or active galactic nuclei \cite{10.1111/j.1365-2966.2012.20879.x}, which were neglected in early simulations.  These effects are difficult to simulate from first principles, due to the vast range of distance scales that need to be considered in structure formation.  The results to date leave doubt as to whether baryonic physics by itself can resolve the small-scale structure puzzles.  Dark matter self-interactions thus continue to be an interesting possible resolution.

The magnitude of the interaction cross section needed is not far below the upper limit implied by Bullet Cluster observations \cite{Randall:2008ppe,Markevitch:2003at,Robertson:2016xjh,Wittman:2017gxn}, $\sigma/m \sim 2$\,cm$^2$/g $\cong 4$\,b/GeV.  These large values are suggestive of the strong interactions of the standard model (SM), making it natural to consider composite dark matter candidates similar to SM nucleons.  An early study
\cite{Cline:2013zca} showed that it is possible to rescale lattice QCD results to make quantitative predictions for the nucleon scattering cross sections in a dark SU(3) model with different confinement scale $\Lambda$ and pion mass $m_\pi$
than in the SM.  Curves in the $m_\pi$-$\Lambda$ parameter space consistent
with the desired value of $\sigma/m$ were identified, assuming a constant (velocity-independent) cross section.

Since that time, it was shown that the cusp-core problem is less
pronounced on the scales of galactic clusters than in individual
galaxies \cite{Newman:2012nv,Newman:2012nw}, suggesting that a unified
solution requires velocity-dependent DM scattering, which would be
weaker for the higher DM velocity dispersions found in clusters
\cite{Kaplinghat:2015aga}.  As was discussed in Refs.\
\cite{Mahbubani:2019pij,Chu:2019awd}, velocity-dependent DM scattering is naturally
accommodated within the effective range formalism for parametrizing
scattering amplitudes.  In this work we extend the analysis of Ref.\
\cite{Cline:2013zca} to include the velocity dependence in a QCD-like
dark sector, which reduces the degeneracy of the favored values of
$m_\pi$ and $\Lambda$ to discrete regions of parameter space.

These conclusions are only relevant if the dark sector consists
primarily of nucleons and not higher-mass dark nuclei, whose
scattering properties would differ from those being calculated here. 
For example, Ref. \cite{Krnjaic:2014xza} (see also Ref.\ \cite{Redi:2018muu}) showed that dark
nucleosynthesis can easily be dominated by high-mass nuclei in a
generic confining dark sector.  Moreover if there are dark electrons,
then nucleon-nucleon scattering might be superseded by atom-atom
scattering.  One must further ensure that dark pions do not dominate
the DM. Hence a further goal of this work is to identify the other
conditions needed to establish that nucleons constitute the dominant
DM component, while fulfilling their QCD-like nature that allows us to
incorporate results from lattice QCD.  We will show that these
requirements put a lower bound on the dark photon mass.

\begin{figure*}[t]
\begin{center}
 \includegraphics[scale=0.50]{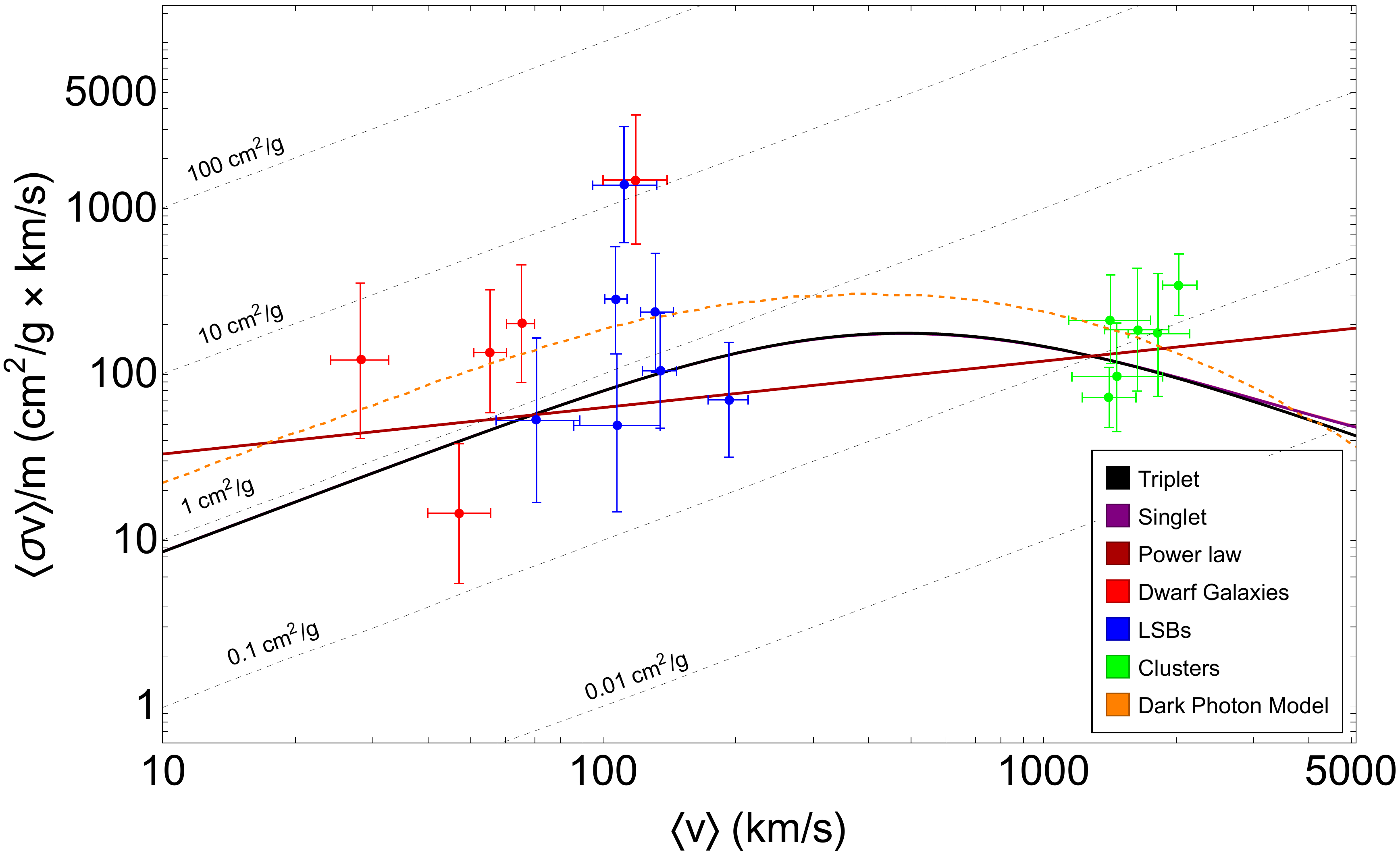}
 \caption{Data points taken from Ref.\ \cite{Kaplinghat:2015aga} for DM self-scattering cross sections $\langle\sigma v\rangle/m$ versus characteristic relative velocity of dark matter particles in galaxies or galactic clusters, and predictions from the dark baryon model (black, present work) and a previously studied DM model with dark photon exchange (orange, Ref.\ \cite{Kaplinghat:2015aga}).  Diagonal lines show $\langle\sigma v\rangle/m$ 
assuming different values of a velocity-independent cross section.
Brown line shows best-fit phenomenological power law.}
 \label{fig:sigma}
\end{center} 
\end{figure*}

\section{Dark nucleon framework}
We assume a dark sector which, like in the SM, has two flavors of light quarks\footnote{and possibly a third heavier (strange) quark, included in some of the lattice simulations relevant for this study \cite{Beane:2005rj}}, whose masses are sufficiently below the confinement scale so that pions can be treated as
pseudo-Goldstone bosons.
This is sufficient for utilizing determinations of nucleon scattering lengths $a_s$
from lattice data, presented in Ref.\ \cite{Chen:2010yt}.  The general form of the $S$-wave scattering amplitude is 
\cite{Chu:2019awd}
\be
    {\cal A}_s = {4\pi\over m_N(-i p - a_s^{-1} + \sfrac12 r_{0,s} p^2 + O(p^4))}\,,
    \label{amp}
\ee 
where $s$ is the spin state of scattering particles (either singlet $s=0$
or triplet $s=1$).
The parameter $r_{0,s}$ is the effective range, which was neglected in Ref.\ \cite{Cline:2013zca}, but {\it a priori} it could be relevant in the present study, since it can affect the velocity-dependence of the cross section at low energy, via 
\be
    \sigma = \sum_{s=0}^1(2s+1)\, {m_N^2|{\cal A}_s|^2\over 16\pi} \,.
    \label{sigma}
\ee

\begin{figure*}[t]
\begin{center}
\centerline{
\includegraphics[scale=0.45]{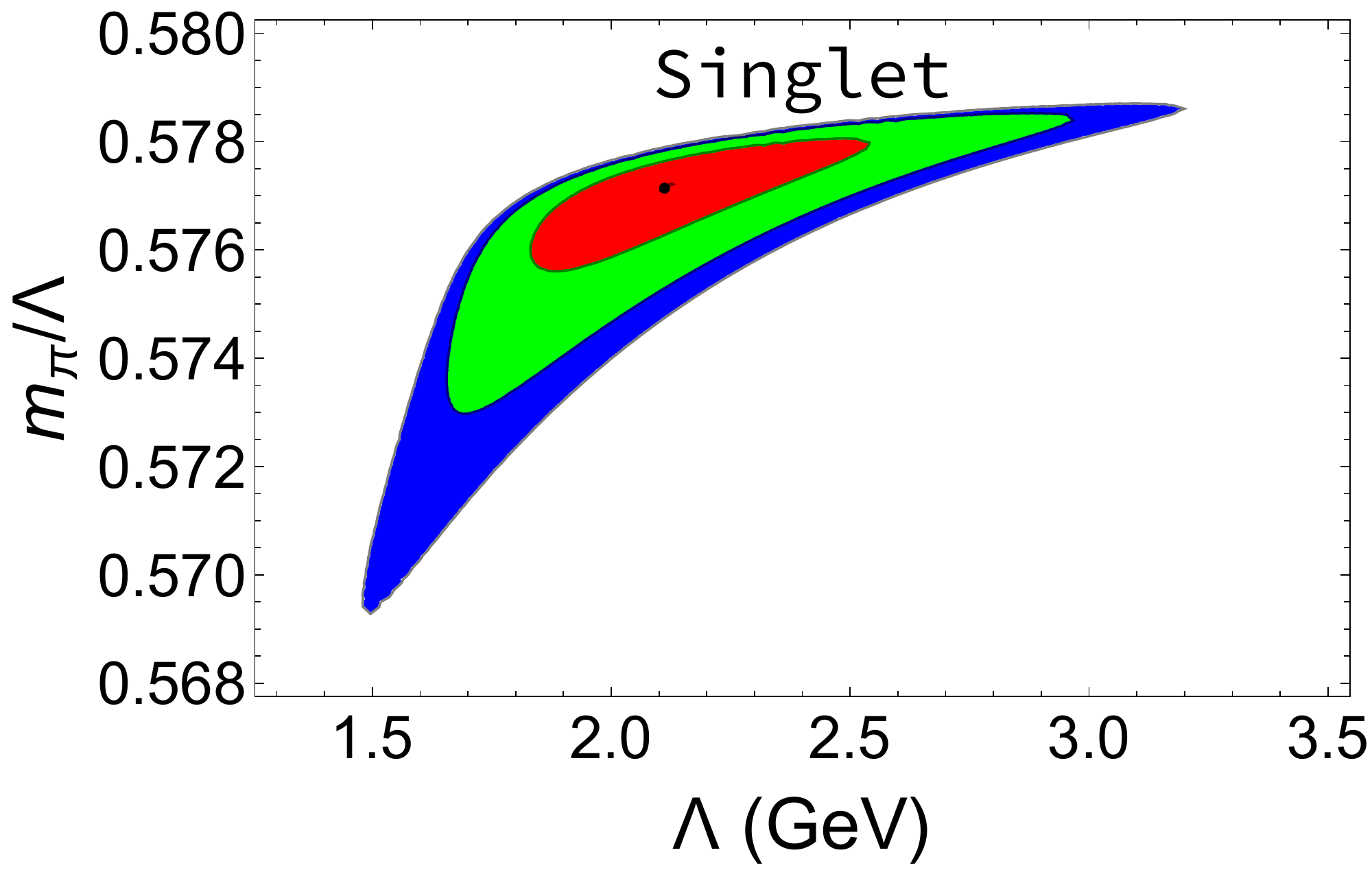} 
\includegraphics[scale=0.45]{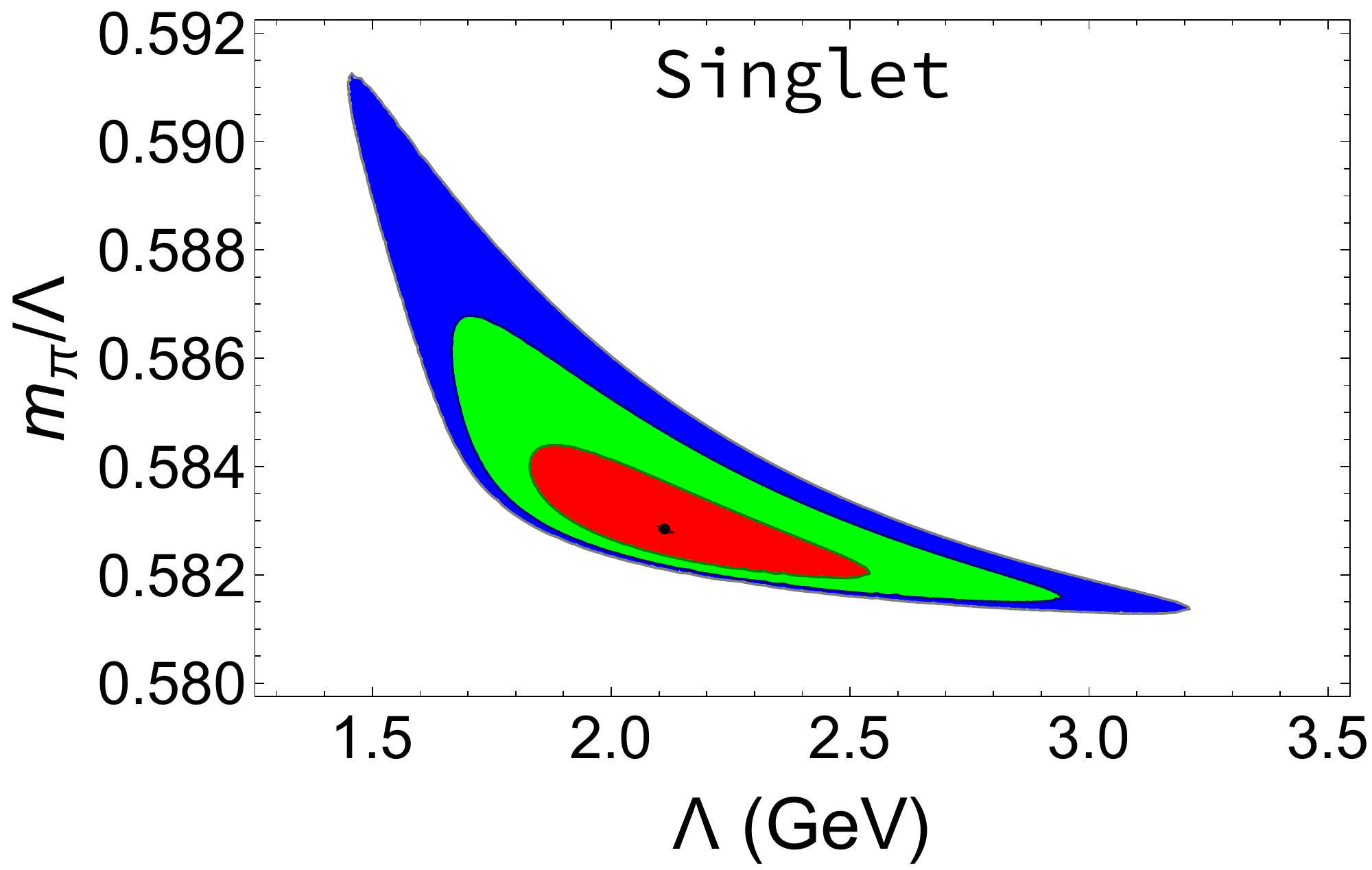}}
\centerline{
\includegraphics[scale=0.45]{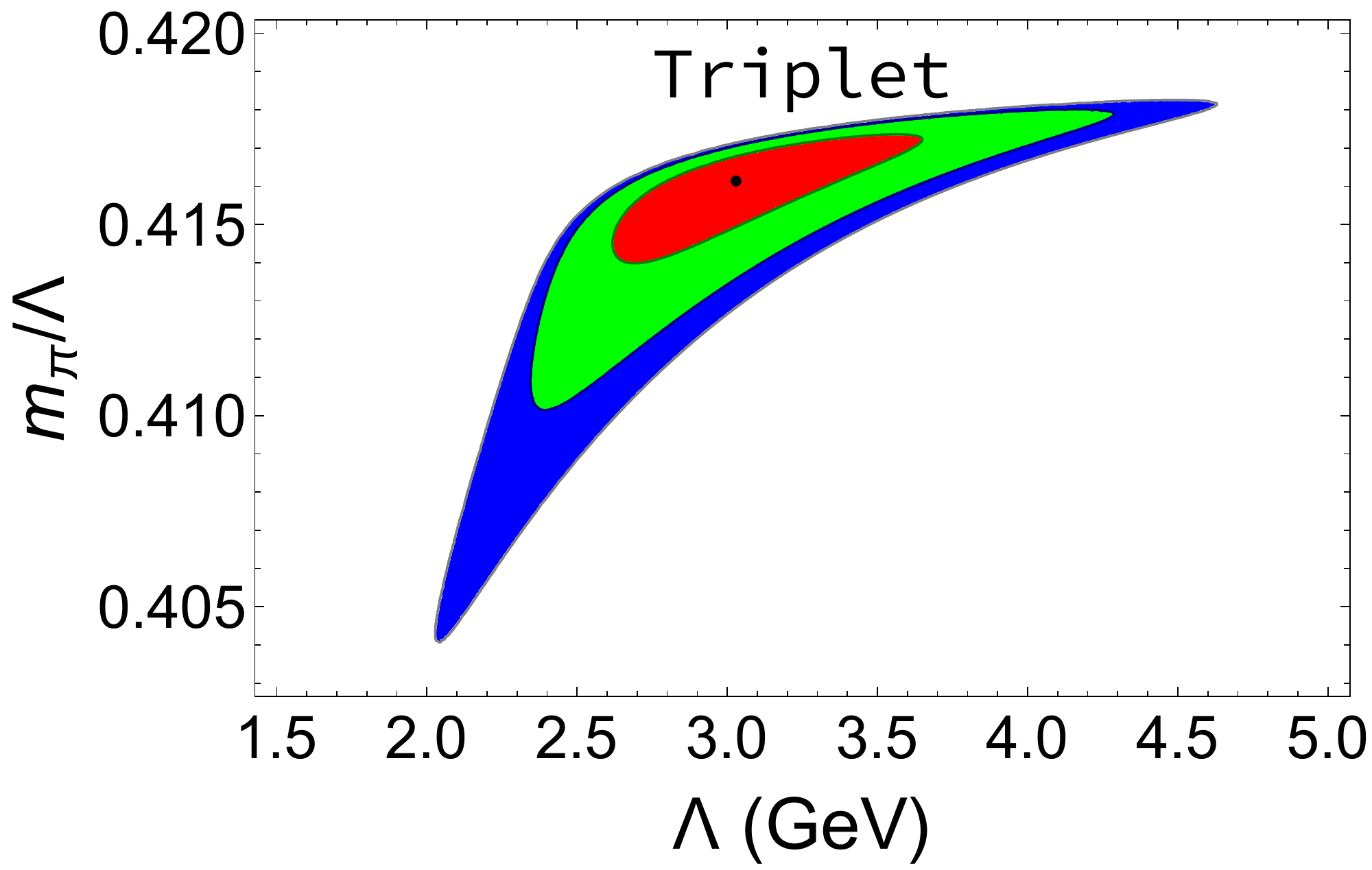}
\includegraphics[scale=0.45]{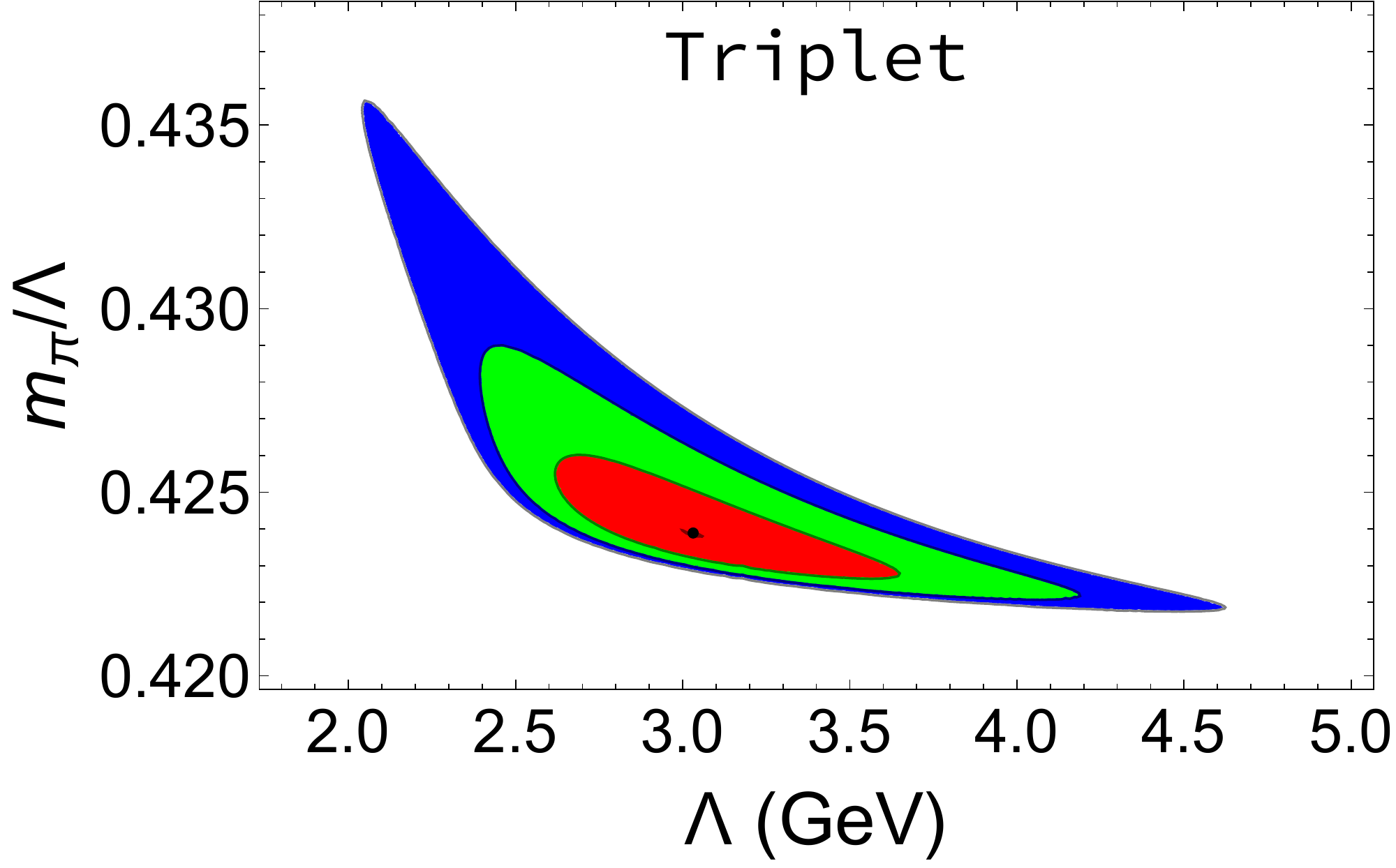} } 
 \caption{Best-fit regions of parameter space for velocity-dependent dark baryon scattering.  Upper (lower) row: scattering dominated by pole
 in singlet (triplet) scattering length,  showing 68\%, 95\%
 and 99\% confidence regions. }
 \label{fig:fits}
\end{center} 
\end{figure*}

 \begin{figure}[t]
\begin{center}
\vskip-1cm
 \includegraphics[scale=0.3]{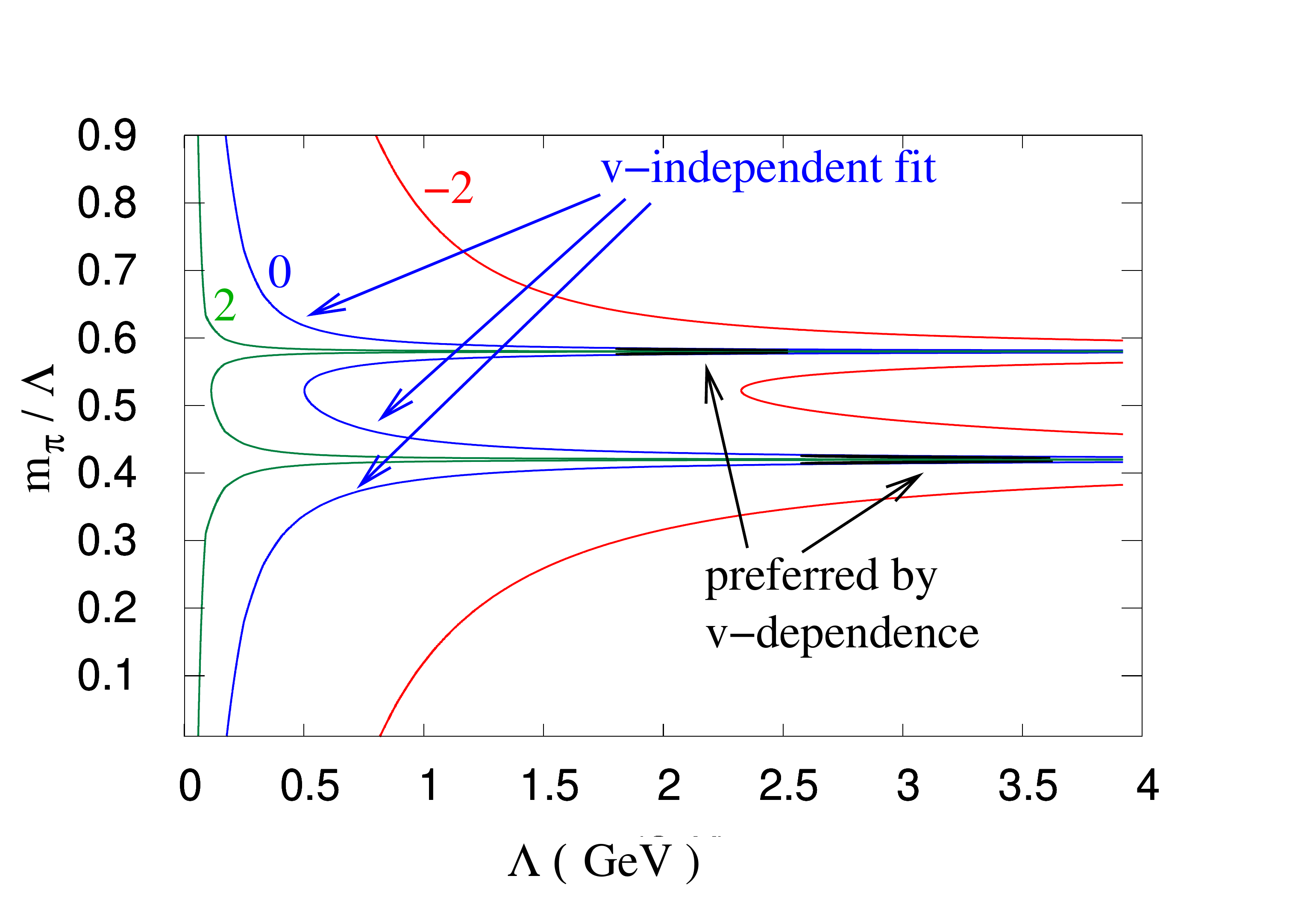}
 \caption{Contours of $\log_{10}[(\sigma/m_N)/(0.6\,{\rm cm}^2/{\rm g})]$
 in the $v\to 0$ limit.
 (colored).  The (blue) ones labeled ``0'' represent the degenerate solutions  for the small-scale structure problems, neglecting velocity dependence of the cross section.  Black regions show the more narrow predictions accounting for
 the velocity dependence, at 68\% confidence.}
 \label{fig:comp}
\end{center} 
\end{figure}

The parameters $a_s$ and $r_{0,s}$ have been fitted to lattice QCD data as a function of the pion mass in Ref.\ \cite{Chen:2010yt}.  By dimensional analysis, these results can be generalized to a QCD-like theory with a different confinement scale, by assuming that $a_s^{-1}$ scales linearly with $\Lambda$, and taking a specific value  for QCD to determine the dimensionless proportionality constant; we adopt $\Lambda = 250\,$MeV.

We find that the scattering lengths,
determined in Ref.\ \cite{Chen:2010yt} using nuclear effective field theory \cite{Beane:2008bt}), can be accurately represented using the simple analytic functions\footnote{This corrects formulas in Ref.\ \cite{Cline:2013zca} that had erroneous digitization of the curves} 
\be
 a_0 = {0.71\, \Lambda^{-1}\over m_\pi/\Lambda-0.58}, \quad a_1 = {0.96\,\Lambda^{-1}\over m_\pi/\Lambda-0.42}\,,
\label{scatt-lengths}
\ee 
given our choice of $\Lambda = 250\,$MeV for QCD.
 In the present study we found no such simple formulas for fitting $r_{0,s}$
 ; instead we digitized the results from Fig.\ 2 of Ref.\ \cite{Chen:2010yt}.  However it turns out that $r_{0,s}\ll a_s$ in the regions of parameter space of interest; hence $r_{0,s}$ can be neglected without appreciably affecting our results for scattering.   

The cross section depends on the relative velocity through $p = m_N v/2 $ in (\ref{amp}),
where $m_N \cong 3.76\,\Lambda$ by normalizing to QCD (in the limit of light quarks whose mass has a negligible effect on $m_N$).  To fit the two free parameters $\Lambda$ and $m_\pi$, we used the inferred values of $\langle\sigma v\rangle/m$
for a selection of galaxies and galactic clusters from Fig.\ 1 of Ref.\ \cite{Kaplinghat:2015aga} to construct a $\chi^2$ function, and minimized it.
This requires performing the phase space average over the DM velocity
distributions, to compute $\langle\sigma v\rangle$.  Assuming a Maxwellian
$e^{-v^2/v_0^2}$ distribution, the averaging can be carried out analytically,
in the approximation of ignoring $r_{0,s}$ in Eq.\ (\ref{amp}), giving
(in $\hbar=c=1$ units)
\bea
    \langle\sigma v\rangle &=& \sum_s (2s+1){ a_s \over m_N} F(b_s);\nn\\
        F(b) &=& \sqrt{\pi/2}\left(2 b + b^{3} e^{b^{2}/2}\,{\rm Ei}(-b^{2}/2)\right)
\eea
where $b_s \equiv 2/(m_N a_s v_0)$, Ei is the exponential integral function, and $v_0 = \sqrt{\pi}\langle v\rangle /2$ in terms of the average DM velocity $\langle v\rangle$.

Fig.\ \ref{fig:sigma} shows the data points and one of our best-fit predictions (black solid line), as well as the prediction from an alternative model involving exchange of a light dark photon (solid orange) \cite{Kaplinghat:2015aga}.  The data points were determined in Ref.\ \cite{Kaplinghat:2015aga} using a semianalytic model of halo profiles calibrated with N-body simulations, assuming a large enough value of $\langle\sigma v\rangle/m$ to explain observed coring of the DM profile in correlation with the velocity dispersion of the system.

We find four best-fit regions of parameter space, shown in Fig.\ \ref{fig:fits}.  These fall on the previously identified best-fit regions 
(blue curves) from Ref.\ \cite{Cline:2013zca},  shown in Fig.\
\ref{fig:comp}, where the velocity-dependence of $\sigma$ was neglected.  The four-fold degeneracy can be understood from 
Eq.\ (\ref{scatt-lengths}): a large enough cross section requires being close to one of the poles of the two scattering lengths, either slightly above or below.  $m_\pi/\Lambda$ must be tuned to one part in $\sim 50$
at the 99\% confidence level (C.L.).  On the other hand, the value of $\Lambda$ is more weakly constrained, $\Lambda \cong 3.0{+1.7\atop-1.0}\,$GeV near the singlet pole or 
$\Lambda \cong 2.1{+1.2\atop-0.6}\,$GeV near the triplet pole.
We find a minimum $\chi^2$ of 22, lower than that of the best-fit dark photon model
shown in Ref.\ \cite{Kaplinghat:2015aga} (which also has two free parameters, the DM and photon masses, with $\alpha'$ held fixed), whose $\chi^2$ is 37. The latter is dominated by contributions from three data points that lie well below the curve, which
cause our best fit curve to be higher.

For comparison, we show the best fit for a hypothetical power law ansatz, $\langle\sigma v\rangle/m = N (v/100\,{\rm km/s})^\alpha$, which also has two free parameters.  It has approximately the same minimum $\chi^2$ as the dark baryon model,
with $N=63$\,cm$^2$/g$\cdot$\,km/s and $\alpha=0.28$.  However, we are not aware of a particle physics model corresponding to this ansatz.

\section{Cosmological consistency constraints}

For these results to be relevant, the dark baryons must be the primary DM components.  In the following, we outline the simplest scenario for ensuring this criterion, and consistency with other constraints.  First, the dark pions (including other pseudoscalar mesons, in the case of three light flavors) and vector mesons must be unstable or subdominant.  Although one could introduce
an analog of weak interactions into the dark sector, a simpler possibility is to require that the $\pi^\pm$ mesons that are charged under a dark U(1)$'$, analogous to electromagnetism, annihilate into dark photons $\gamma'$ to a low abundance.\footnote{We assume that the mechanism that generates the dark baryon asymmetry does not also create a dark pion asymmetry.
The case of symmetric dark baryons created by freeze-in has been
considered in Ref.\ \cite{Garani:2021zrr}. }  This implies \cite{Steigman:2012nb}
\be
    \langle\sigma v\rangle_{\pi^+\pi^-\to\gamma'\gamma'} \sim {g^4\over \pi m_\pi^2} \gg 
     \langle\sigma v\rangle_0 \equiv 3\times 10^{-26}{\rm cm^3/s}\,,
\label{piann}
\ee
where $g$ is the U(1)$'$ coupling.   Taking $m_\pi\sim 1$\,GeV, we find the modest requirement $g\gg 10^{-4}$.  For example if  $\alpha' = g^2/4\pi = 10^{-3}$, 
dark mesons would constitute only $10^{-5}$ of the DM.  Similar conclusions hold for dark $K^0$ and $K^+$ in the three-flavor case (assuming the strange quark mass is not much greater than $m_\pi$).  

Even a small residual rate of dark 
$\pi^\pm\to\gamma'\gamma'$ annihilations may have
observable effects at late times, since the $\gamma'$ decays can
distort the cosmic microwave background (CMB) \cite{Slatyer:2015jla}, 21-cm signal
\cite{Liu:2018uzy}, and big bang nucleosynthesis (BBN) \cite{Depta:2019lbe}.  For GeV-scale DM,
the CMB constrains $\langle\sigma v\rangle\lesssim 10^{-27}$\,cm$^3$/s,
which is not far below the fiducial cross section (\ref{piann}).  A
subdominant DM component like our dark $\pi^\pm$ would thus be
unconstrained.

Considering the neutral $\pi^0$ and $\eta$, their decays into
$\gamma'\gamma'$ will be fast, so long as $m_\gamma' < m_\pi/2 \sim
500\,$MeV. Similarly, dark glueballs have a mass of order $7\Lambda$
in QCD, which is far above the threshold for decay into two or three
mesons in our scenario.  They are therefore short-lived and pose no
risk as DM relics. One expects vector mesons to decay to pions, like in QCD.

If the dark photon is still massless when the dark proton asymmetry is generated in the early universe, there should be a compensating dark electron ($e'$) asymmetry to ensure U(1)$'$ charge neutrality.  There is then the danger that dark H atoms constitute a large fraction of the DM, and self-interact more strongly than nucleons.  This can be avoided if the dark photon mass is large enough to inhibit the formation of bound states \cite{Petraki:2016cnz,Cline:2021itd}:
\be
    m_{\gamma'} \gtrsim \alpha'{ m_{e'} m_N\over m_{e'}+m_N}\,,
    \label{dabound}
\ee
where $\alpha' = g^2/4\pi$.  Since this depends on the dark electron mass, it does not give an independent constraint on $m_{\gamma'}$.  In fact, for the
preferred regions of $m_N$ and the small value $\alpha'\sim 10^{-3}$ adopted below, the constraint (\ref{dabound})
is satisfied for any $m_{e'}$.

Next we consider how to prevent dark nucleons from binding
significantly into larger nuclei.  It is interesting that the
criterion of being close to the pole of the scattering length implies
that there is no bound deuteron $d$ in the case where $m_\pi/\Lambda =
0.42 + \epsilon$, which provides an obstruction to nucleosynthesis
from proceeding through the reaction $n+p \to d + \gamma'$.  More
generally, one can block dark BBN (big bang nucleosynthesis) by taking
$m_{\gamma'}$ to exceed the binding energy of the deuteron, or of the
spin singlet dinucleon state in the isotriplet channel $N_i + N_j \to
D_{k} + \gamma'$, where $N_i$ are the nucleon states and $D_{k}$ is
the dinucleon isotriplet.  

The binding energies of $d$ and $D_{k}$ as determined by lattice QCD
studies, combined with effective field theory, are given as a function
of $m_\pi$ in Ref.\ \cite{Chen:2010yt}.  In fact they are given in
terms of the effective range parameters by $E_{b,s} = (m_N
r_{0,s}^2)^{-1}(1 - \sqrt{1 -2 r_{0,s}/a_s})^2 \cong (m_N a_s)^{-2}$ for
$|r_{0,s}|\ll |a_s|$, and binding can only
occur for the ranges of $m_\pi/\Lambda$ where $a_s>0$ and $0 <
2r_{0,s}/a_s < 1$.  (These ranges are such that there is no overlap
between the singlet and triplet: at most one channel can have a bound
state.) We find that the binding energies are of order $0.01$\,MeV.
Therefore if
$m_{\gamma'}\gtrsim 0.01\,$MeV, dark BBN is generally inhibited.
The dark pions
themselves are much heavier than this scale, and therefore cannot
serve as a light mediator for carrying away binding energy to form $d$
or $D_k$.  Aside from kinematically blocking $d$ or $D_k$ production, a
sufficiently small coupling $\alpha'$ will impede these processes,
as was quantified in Ref.\ \cite{Redi:2018muu}.

Beyond dark BBN considerations, the fusion of dark nucleons
into dinucleon bound states plus dark photon is constrained by
indirect signals, notably the effect of energy injection from dark
photon decays into the cosmic microwave background (CMB) 
\cite{Mahbubani:2019pij}.  By blocking bound state production we also
satisfy these constraints.

\begin{figure}[t]
\begin{center}
 \includegraphics[scale=0.25]{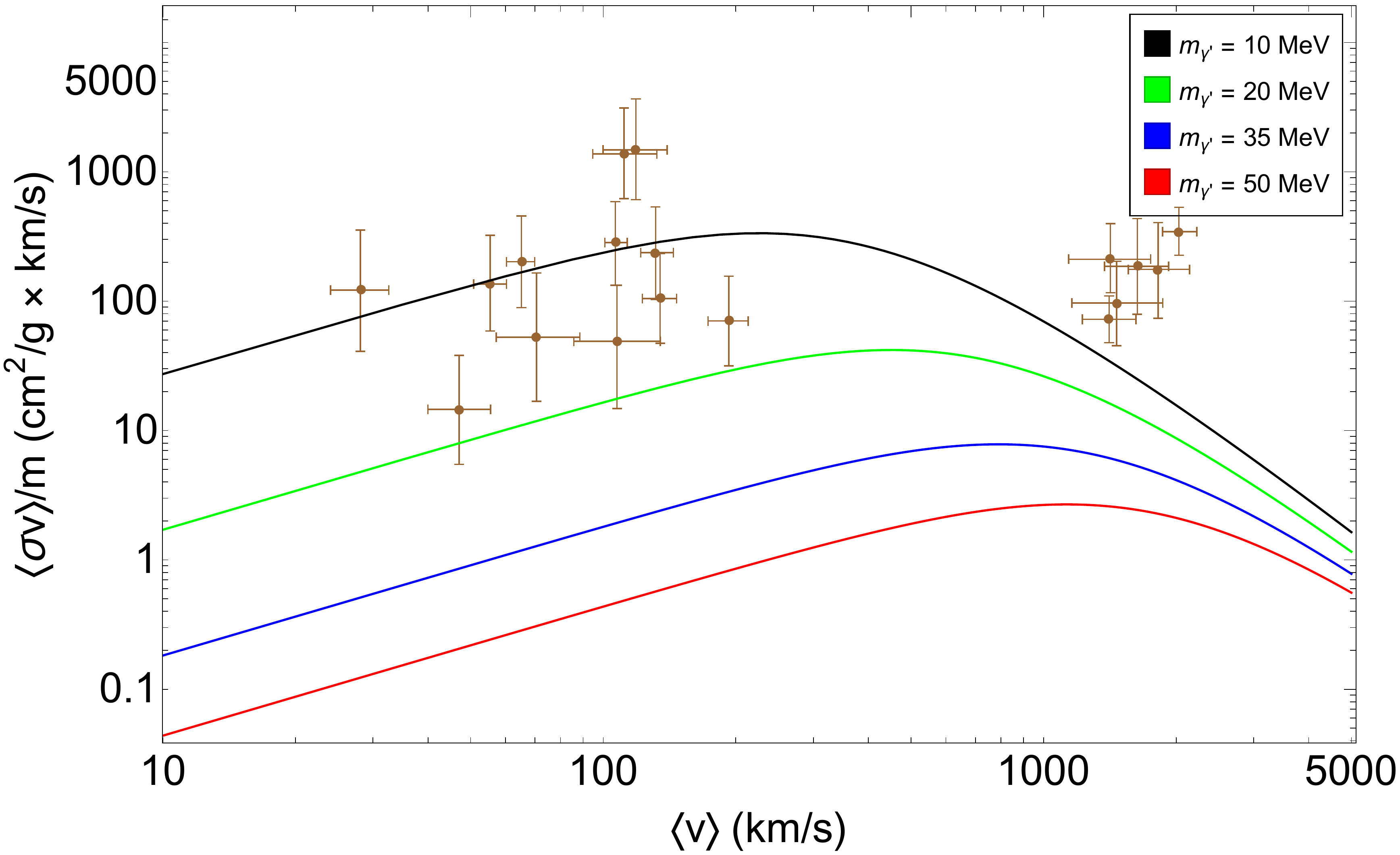}
 \caption{Perturbative predictions of the dark-photon mediated contribution to scattering of dark photons, for $\alpha'=10^{-3}$ and $m_{\gamma'}=10,\,20,\,35,\,50$\,MeV, from top to bottom. The DM mass is taken to be 10\,GeV. Experimental data points from Fig.\ \ref{fig:sigma} are also plotted.}
 \label{fig:svm}
\end{center} 
\end{figure}

A further requirement is that nucleon self-interactions mediated by dark photon exchange should be subdominant to the strong interactions.  This puts a stronger constraint on $m_{\gamma'}$ than does dark BBN.  The scattering can be computed nonrelativistically using the Yukawa potential $\alpha' e^{-m_{\gamma'}r}/r$.  Depending on the values of the parameters $R = (m_N v/m_{\gamma'})^2$ and $Q=\alpha' m_N/m_{\gamma'}$, where $v$ is the relative velocity, the cross section can be Sommerfeld enhanced.  If both $R,Q \ll 1$, the Born approximation is valid, and the momentum transfer cross section is given by \cite{Tulin:2013teo}
\be
    \sigma_T = {8\pi\alpha'^2\over m_N^2 v^4}\left[\ln(1+R) - {1\over 1 + R^{-1}}\right]\,.
    \label{sigmaT}
\ee
For $\alpha' =10^{-3}$ and $m_N\sim 10\,$GeV, this approximation is adequate for $m_{\gamma'}\gg 10\,$MeV; otherwise a numerical solution of the Schr\"odinger equation going beyond perturbation theory is required.  For simplicity, we use Eq.\ (\ref{sigmaT}) to estimate the lower bound on 
$m_{\gamma'}$ (taking benchmark value $\alpha' =10^{-3}$), and cross-check our conclusion against the numerical results of Ref.\ \cite{Tulin:2013teo}.
The $v$-dependence of $\sigma_T v/m$ is plotted in Fig.\ \ref{fig:svm} for
several values of $m_{\gamma'}$, and assuming $m=10\,$GeV,  compatible with models in Table \ref{tab1}.   Comparison with the data suggests that
if $m_{\gamma'} \gtrsim 35\,$MeV, this contribution to dark nucleon scattering is unimportant.  This is corroborated by Fig.\ 6 of Ref.\ \cite{Tulin:2013teo}, which takes account of nonperturbative effects.

Lastly, the dark photons must be unstable, to avoid overclosing the Universe
\cite{Kaplinghat:2013yxa}.  This is usually accomplished by introducing a kinetic mixing
Lagrangian $-(\epsilon/2)F^{\mu\nu}F'_{\mu\nu}$ with the SM hypercharge;
then $A'_\mu$ couples to $\epsilon e$ times the SM currents of charged
particles, allowing $\gamma'\to e^+ e^-$ decays. For $m_{\gamma'} \sim 35\,$MeV, there are windows of allowed $\epsilon$ in the vicinity of $10^{-4}$ \cite{Fabbrichesi:2020wbt} and $\lesssim 10^{-10}$, where the latter is the bound from supernova 1987A \cite{Chang:2016ntp,Li:2020roy}.  The former value $\sim 10^{-4}$ is excluded in the present model beam dump
experiments (see Fig.\ 3.4 of \cite{Fabbrichesi:2020wbt}).  The cross section for $pp'\to pp'$ elastic scattering is given by 
\be
    \sigma_p \cong {(\epsilon e g m_p)^2\over \pi m_{\gamma'}^4} \lesssim 10^{-45}\,{\rm cm}^2\,,
\ee 
where $m_p$ is the ordinary proton mass and the upper limit
is from Refs.\  \cite{XENON:2018voc,PandaX-II:2018xpz}.  Taking
$\epsilon = 10^{-10}$ and $\alpha'=10^{-3}$, we find $\sigma_p = 2\times 10^{-46}\,$cm$^2$.  Even though this is only a factor of 5 less than the current limit, it is below the sensitivity of currently planned experiments if the dark nucleon  mass is $m_N\lesssim 9\,$GeV \cite{Billard:2021uyg},
which is compatible with the allowed regions from our scattering fits.  However for somewhat larger masses $\gtrsim 10\,$GeV, also allowed, this cross section would be observable through direct detection, due to the quickly rising sensitivity as the recoil energy increases.\footnote{
Alternatively, one could turn off the kinetic mixing and allow $\gamma'$ to decay into dark radiation ({\it e.g.}, massless fermions), thereby circumventing all of these bounds.}

\section{Conclusions}

In summary, we find that the strong interactions of a QCD-like dark nucleon sector could have the desired velocity-dependent self-scattering cross section for solving the cusp-core problem of dark matter halos, on scales from dwarf galaxies to galactic clusters.  At the same time, a variety of
self-consistency requirements must be satisfied for this scenario to be realistic, necessitating the existence of dark photons and electrons in addition to the baryons.  A portal between the dark sector and the SM is needed to deplete dark pions, through annihilation or decay, which we took to be kinetic mixing of $\gamma'$ with hypercharge.  The $\gamma'$ mass should be between $\sim 35$\,MeV (to sufficiently deplete dark pions via annihilation) and $m_\pi/2$ (to allow decay of neutral pions into $\gamma'\gamma'$).  The $\gamma'$ must decay into SM $e^+ e^-$ or dark radiation; in the former case, kinetic mixing at the level of $\epsilon\sim 10^{-10}$ is needed to avoid supernova and nucleosynthesis constraints on $\gamma'$.  The model predicts a level of dark proton-proton scattering that could be detectable in direct searches, depending on $m_N$. 
Benchmark values of allowed parameters are summarized in Table \ref{tab1}.

\begin{table}[t]\centering
\setlength\tabcolsep{4pt}
\def\arraystretch{1.2}
\begin{tabular}{|c|c|c|c|c|c|c|}
\hline
${m_\pi/\Lambda}$ & $\Lambda$ & $m_N$ & $m_\pi$ & $m_{\gamma'}$  & $\alpha'$ & $\epsilon$\\
\hline
$\!\!\!\!_{\phantom{|_|}}{0.58\atop 0.42}^{\phantom{|^|}}\!\!\!\!$ & ${1.5{\rm-}3.3\atop 2.0{\rm-}4.7}$ & ${5.6{\rm-}12\atop 7.5{\rm-}18}$ & ${0.87{\rm-}1.9\atop
0.84{\rm-}2.0}$& $\gtrsim 0.035$ & $10^{-3}$ &  $10^{-10}$\\
\hline
\end{tabular}
\caption{Values of model parameters in benchmark models.  $\Lambda$, $m_N$ $m_\pi$ and $m_{\gamma'}$ are in GeV/$c^2$ units.}
\label{tab1}
\end{table}

One peculiarity of this scenario is that the ratio $m_\pi/\Lambda$ is tuned at the level of 2\% to give a large enough scattering length, in either the singlet or triplet channel, which has the consequence of making the deuteron or dinucleon being very close being a zero-energy (quasi)bound state.  Curiously, a similar coincidence occurs in the SM QCD sector, where $m_\pi$ is within 5\% of the boundary for no deuteron bound state.  This is the origin of the famous deuterium bottleneck of BBN.

In our study we chose a particular EFT matching to rather old lattice QCD data in Ref.\ \cite{Chen:2010yt}.  However there are other EFT choices that one could make, reviewed in Ref.\ \cite{Hammer:2019poc}, and newer data.  It has been argued
in Ref.\ \cite{Nicholson:2021zwi} that there remains a large degree of theoretical uncertainty
in  EFT extrapolations to lower pion masses than can currently be
achieved on the lattice.  We have not attempted to quantify these
uncertainties, but we do not expect them to affect the qualitative
features of our conclusions.
Namely, all such studies predict scattering lengths of the general form
in Eq.\ (\ref{scatt-lengths}).  Hence the ultimate values of $\Lambda$ and $m_\pi/\Lambda$ needed to fit cosmological data may differ by factors of order 1 from those we have found, but the predicted shape of the $\langle\sigma v\rangle/m$ curve shown in Fig.\ \ref{fig:sigma} should remain substantially unchanged.  In any case, the best fit values of the dark QCD parameters will also depend on the light quark mass spectrum, which is an additional model-building input.

We have not tried to address how the needed asymmetry between dark baryons and antibaryons could be generated.  Ref.\ \cite{Frandsen:2011kt} proposed a model of dark baryons whose relic density is naturally achieved through dynamical electroweak symmetry breaking. It was recently shown that
a dark sector with confining SU(2) interactions can be unified with SU(3)$_{\rm color}$ to provide a UV-complete explanation of simultaneous baryogenesis in both sectors \cite{Murgui:2021eqf}.  In this framework, the relative closeness of the asymmetry in the two sectors, as observed in CMB data, is a consequence of a mild hierarchy $\Lambda/\Lambda_{QCD}\sim 6$ between the confinement scales.  Our model
has a similar hierarchy.  It might be interesting to investigate whether unification of SU$(3)_{\rm dark}\times$SU(3)$_{QCD}$ to SU(6) could 
 provide a more complete framework for self-interacting 
dark baryons of the kind we have considered.

\bigskip {\bf Acknowledgments.}   We thank G.\ Alonso-\'Alvarez, K.\
Moorthy, M.\ Redi, J.-S.\ Roux and A.\ Tesi for helpful discussions. We thank anonymous referees for very useful suggestions. This research
was supported by the Natural Sciences and Engineering Research Council
(NSERC) of Canada.

\bibliography{ref}
\bibliographystyle{utphys}

\end{document}